\documentclass[twocolumn,showkeys,showpacs,aps]{revtex4}

\usepackage{amssymb}
\usepackage{dcolumn}
\usepackage{amsmath}

\begin{document}
\title{Twistor variational principle for null strings}
\author{Kost' Ilyenko}
 \thanks{Also at Faculty of Physics, Kharkiv National University.}%
 \email{kost@ire.kharkov.ua}
\affiliation{Department of Theoretical Physics, Institute for
             Radiophysics and Electronics of NASU, 12 Acad.
             Proskura Street, Kharkiv 61085, Ukraine}%
\date{\today}%
\begin{abstract}
I present a twistor action functional for null 2-surfaces (null
strings) in 4D~Minkowski spacetime. The proposed formulation is
reparametrization  invariant and free of algebraic and
differential constraints. Proposed approach results in derivation
of evolution equations for the null strings. It is shown that
non-geodesic null strings are contained in the presented
formalism. A discussion of the problem of minimality for
2-surfaces with degenerate induced metric is given. I also
speculate on the possible description of strings (time-like
2-surfaces) and conventional (space-like) 2-surfaces.%
\end{abstract}
\maketitle

\section{INTRODUCTION}

The notion of null 2-surface was put forward by A.~Schild as a
tensionless (null) string, i.e. a 2-dimensional ruled by a
one-parameter family of light-like geodesics submanifold of
4D~Minkowski or curved spacetime. The induced metric on such
submanifolds is degenerate:
\begin{equation}
\dot{x}^2\acute{x}^2 - (\dot{x}\acute{x})^2 = 0.%
\label{NullDet}
\end{equation}
Here $x^a (\tau,\sigma)$ is the world-sheet coordinate, the dots
and primes denote differentiation with respect to the parameters
$\tau$ and $\sigma$, $\dot{x}\acute{x}$ stands for
$\dot{x}{}^a\acute{x}_a$. It should be emphasized that the null
property (\ref{NullDet}) is manifestly reparametrization invariant
while the original Schild's variational principle lacked this
feature~\cite{Schild}.

A different approach was developed in the
articles~\cite{Stachel,Gusev-Zheltukhin}. In the
paper~\cite{Stachel} J.~Stachel chose a null bivector (2-form
obeying algebraic constraints
\begin{equation}
p_{ab}(x){}^*p^{ab}(x)=p_{ab}(x)p^{ab}(x)=0,
\end{equation}
the star stands for duality operation) as the Lagrangian density
and showed that Schild's null strings can be described in this
way. He also imposed the integrability condition
\begin{equation}
p_{a[b}(x)\nabla_{c}p_{de]}(x)=0%
\label{DiffCon}
\end{equation}
for null bivectors (square brackets denote
skew-sym\-met\-ri\-za\-tion, $\nabla_a$ stands for the covariant
de\-ri\-va\-ti\-ve). It stems from the requirements of
differential geometry and was adopted from the
book~\cite{Schouten}. Recently, O.E.~Gusev and
A.A.~Zheltukhin~\cite{Gusev-Zheltukhin} have solved the algebraic
constraints in the physical dimensions of Minkowski spacetime
using a fundamental result of spinor calculus of
Cartan-Penrose~\cite[Vol. 1]{SST}. According to this result a null
bivector can be represented with the aid of a 2-component spinor
and the action principle takes the form
\begin{equation}
S = \int (\bar{\pi}_{A}\bar{\pi}_{B}\epsilon_{A'B'}\! +
\pi_{A'}\pi_{B'}\epsilon_{AB}) dx^{AA'}\!\!\! \wedge
dx^{BB'}\!\!\!.%
\label{GusZhelt}
\end{equation}
Using this action, they proved the null property~(\ref{NullDet})
of the resulting 2-surfaces and treated the case of geodesic null
strings.

D.V.~Volkov and his co-authors were interested in applications of
twistors to supersymmetric  objects (superbranes) and laid the
foundations of the superembedding approach
\cite{Volkov:1988vf,Sorokin:1989zi,Sorokin:1989nj}.

In the present contribution I show that the variational principle
(\ref{GusZhelt}) admits a natural twistor generalization.
Additionally to the solutions already found
in~\cite{Gusev-Zheltukhin}, the corresponding Euler-Lagrange
equations contain as their solutions generic, i.e. not necessarily
ruled by null geodesics of the ambient spacetime and referred to
here as non-geodesic, null 2-surfaces (null strings). I derive a
non-linear evolution equation governing the propagation of a
generic non-geodesic null string. A heuristic argument in favour
of defining the minimal null 2-surfaces as ruled by null geodesics
of the ambient spacetime is presented. I also draw attention of
the reader to the possibility and advantages of using the proposed
formalism for expressing variational principles for strings
(time-like 2-surfaces) and conventional (space-like) 2-surfaces of
4D~Minkowski spacetime (cf. \cite{Gusev-Zheltukhin}).

\section{ACTION FUNCTIONAL}

By definition, the null bivector, $p_{ab}$, obeys the pair of
above stated algebraic constraints. This fact leads to the
existence of a pair of 1-forms with components $u_a(x)$ and
$v_b(x)$ such that $p_{ab} = (1/2!)(u_a v_b - u_b v_a)\equiv
u_{[a}v_{b]}$. Moreover, it follows that without loss of
generality one has $u_a u^a = u_a v^a = 0$, $v_a v^a < 0$, and the
particular value of the Lorentz norm of $v_a$ is irrelevant for
the variational problem in question~\cite{Schouten}.

Let the spinor fields $\bar{\pi}_A(x)$\hspace{1ex} and
$\bar{\eta}_A(x)$\hspace{1ex} to con\-s\-ti\-tu\-te a normalized
Newman-Penrose dyad ($\bar{\pi}_A \bar{\eta}^A = 1$) in
4D~Minkowski spacetime and additionally assume the spinor field
$\bar{\pi}_A$ to be chosen in such a way as to represent the
coincident principal null directions of the null bivector
$p_{ab}$. Then, one can write $u_a = \bar{\pi}_A\pi_{A'}$ and $v_a
= \bar{\pi}_A\eta_{A'} + \pi_{A'}\bar{\eta}_A$. This
representation is, up to an overall functional multiplier, the
general solution of the algebraic constraints for the null
bivector $p_{ab}(x)$ (see, for example, \cite{SST}). To switch
between vector and spinor indices the conventions of \cite{SST}
are used.

If one assembles a pair of 1-forms $u_a dx^a = \bar{\pi}_A
\pi_{A'} dx^{AA'}$ and $v_a dx^a = (\bar{\pi}_A \eta_{A'} +
\pi_{A'}\bar{\eta}_A)dx^{AA'}$, introduces the null twistors
$Z^\alpha \equiv (\bar{\omega}^A , \pi_{A'})$ and $W^\alpha \equiv
(\bar{\xi}^A , \eta_{A'})$, expresses the 1-forms in terms of the
null twistors and substitutes the results in the formula for
$p_{ab}dx^{a}\wedge dx^{b}$ then (s)he obtains the following
twistor variational principle for a null string in 4D~Minkowski
spacetime:
\begin{equation}
S = \int \overline{Z}_\alpha dZ^\alpha \wedge \left(
\overline{Z}_\beta dW^\beta + \overline{W}_\beta dZ^\beta
\right).%
\label{Action}
\end{equation}
Here the spinor fields $\bar{\omega}{}^A$ and $\bar{\xi}{}^A$ are
given by the usual definitions: $\bar{\omega}^A =
ix^{AA'}\pi_{A'}$ and $\bar{\xi}^A = ix^{AA'}\eta_{A'}$.
$\overline{Z}_\alpha$ and $\overline{W}_\alpha$ are the conjugate
null twistors. The null property of the twistors $Z^\alpha$ and
$W^\alpha$ correspond to the Hermitian property of $x^{AA'}$ and
leads to the identities $\overline{Z}_\alpha Z^\alpha =
\overline{W}_\alpha W^\alpha = \overline{Z}_\alpha W^\alpha =
\overline{W}_\alpha Z^\alpha = 0$. It also reflects the reality
conditions imposed on the points of 4D~Minkowski spacetime. The
2-form in (\ref{Action}) is understood to be restricted to a
2-dimensional submanifold of 4D~Minkowski spacetime parametrized
by $\tau$ and $\sigma$.

The Lagrangian density of the twistor action functional
(\ref{Action}) is multiplied by the factor $q^2$ under the gauge
transformations of the form
\begin{equation}
Z^\alpha \rightarrow qZ^\alpha ,\,\,\,\, W^\alpha \rightarrow
q^{-1}W^\alpha + pZ^\alpha .
\end{equation}
Here $q(\tau ,\sigma)$ is a nowhere vanishing real-valued function
and $p(\tau,\sigma)$ is an arbitrary com\-plex-valued function.
This is an admissible freedom for a differential form representing
a surface~\cite{Schouten}. It gives rise to invariance of the
Euler-Lagrange equations under the above mentioned
transformations. The invariance corresponds to the possibility of
rescaling with real multiples of the extent of the null direction
tangent to the null string world-sheet $\bar{\pi}^A \rightarrow
q\bar{\pi}^A$ and to an addition of real multiples of the null
direction to the space-like direction tangent to the world-sheet
$\bar{\eta}^A \rightarrow q^{-1}\bar{\eta}^A + p\bar{\pi}^A$. Thus
the gauge freedom of the null string world-sheet comprises the
null- and boost-rotations.

\section{EVOLUTION EQUATIONS}

The Euler-Lagrange equations for the variational principle
(\ref{Action}) were obtained in the author's doctoral thesis
\cite{Disser}. After some tedious but straightforward algebra they
lead to the property (\ref{NullDet}) (cf.
\cite{Gusev-Zheltukhin}). In addition, one can show that the
differential constraint does not introduce new equations to those
obtained by variational procedure from the action (\ref{Action}).
The latter can be proved by expressing the differential constraint
in terms of spinor fields $\bar{\pi}{}^A$ and  $\bar{\eta}{}^A$
(see \cite{Disser}).

It follows that the twistor action functional (\ref{Action})
describes the null string as a 2-dimensional submanifold of
4D~Minkowski spacetime with the degenerate induced metric. It is
also remarkable that this formulation is reparametrization
invariant and free of additional algebraic and differential
constraints present in the spacetime description \cite{Stachel}
and of the auxiliary world-sheet quantities artificially
introduced in an earlier formulation by I.A.~Bandos and
A.A.~Zheltukhin \cite{Bandos-Zheltukhin2}.

One can choose orthogonal gauge for the world-sheet of a null
string so that $\dot{x}{}^2=0$ and (\ref{NullDet}) implies
$\dot{x}\acute{x}=0$. In spinor terms this means that
$\dot{x}{}^{AA'}=r\bar{\pi}{}^A\pi{}^{A'}$ and
$\acute{x}{}^{AA'}=i(\zeta\bar{\eta}^A\pi^{A'} -
\bar{\zeta}\bar{\pi}^A\eta^{A'})$, where $r(\tau,\sigma)$ is the
real-valued flagpole extent and the function $\zeta(\tau,\sigma)$
is complex valued. The equations of motion for the variational
principle (\ref{Action}) also imply that
$\dot{\bar{\pi}}{}^A(\zeta-\bar{\zeta})=0$. This means either that
$\dot{\bar{\pi}}{}^A=0$ or the function $\zeta$ is real-valued.

The first opportunity was spotted by O.E.~Gusev and
A.A.~Zheltukhin in the article \cite{Gusev-Zheltukhin}. It
immediately leads to the equation $\ddot{x}{}^a \propto
\dot{x}{}^a$, which states that integral curves of the vector
field $\dot{x}{}^a$ are (null) geodesics in the sense of the
ambient Minkowski spacetime. If the affine parametrization for the
null geodesics is chosen then one can write an evolution equation
for \textit{geodesic null strings} in the form:
\begin{equation}\label{EvGeod}
\ddot{x}{}^a=0.
\end{equation}

The second opportunity allows one to take the so-called natural
parametrization, $r^{-1}(\tau,\sigma)=\kappa\equiv
\bar{\pi}_A\bar{\pi}{}^B\pi{}^{B'}$ $\nabla_{BB'}\bar{\pi}{}^A$
(for the geodesic case the spin-coefficient $\kappa$ vanishes),
and results in the following complete set of equations of motion
for non-geodesic null strings:
\begin{eqnarray}
\dot{x}^{AA'}\!\! = \kappa^{-1}\bar{\pi}^A\pi^{A'}\!\!,\,\,
\acute{x}^{AA'}\!\!\!\!\!\!&=&\!\!\!\! i\zeta
(\bar{\pi}{}^A\eta^{A'} - \bar{\eta}{}^A\pi^{A'}), \nonumber \\
\kappa^{-1}(\bar{\pi}^A\acute{\bar{\pi}}_A +
\pi^{A'}\acute{\pi}_{A'})\!\!\!\!&=&\!\!\!\!
2i\zeta(\pi^{A'}\dot{\eta}_{A'} -
\bar{\pi}^A\dot{\bar{\eta}}_A), \nonumber \\
\bar{\pi}^A\dot{\bar{\pi}}_A -
\pi^{A'}\dot{\pi}_{A'}\!\!\!\!&=&\!\!\!\!0.%
\label{RMEqns}
\end{eqnarray}
The spin-coefficient $\kappa$ reflects existence of interaction
between the null string and external fields. Such interactions
preserve the null character of the string world-sheet but violate
its geodesic property. For an example see the article
\cite{Ilienko2}. It should be pointed out that the set of
equations (\ref{RMEqns}) coincides with the system derived in that
paper for the non-geodesic (interacting) null string. This proves
an equivalence on the classical level of the both approaches.

In the natural parametrization one finds the next identities
$\dot{\bar{\pi}}_A\bar{\pi}{}^A=1$, and it follows that
$\bar{\eta}{}^A=-\dot{\bar{\pi}}{}^A$. These results can be used
to show that there exists a non-linear evolution equation for
\textit{non-geodesic null strings}
\begin{equation}
\ddot{x}^2\big[\acute{x}{}^2x\!\!\dot{}\,\acute{}\,{}^a -
(\acute{x}x\!\!\dot{}\,\acute{}\,)\acute{x}{}^a\big] -
\acute{x}^2(\ddot{x}x\!\!\dot{}\,\dot{}\,\acute{}\,)\dot{x}{}^a =
0.%
\label{EvEqn}
\end{equation}
The non-geodesic null string evolution equation is invariant under
a subgroup of diffeomorphisms of the null string world-sheet which
preserves the orthogonal gauge.

\section{MINIMAL NULL 2-SURFACES}

The results of Hughston and Shaw~\cite{Hughston-Shaw} on the
connection between non-interacting (free) strings and minimal
time-like 2-surfaces in 4D~Minkowski spacetime provide an impetus
for attempts of finding a similar correspondence between geodesic
(i.e. non-interacting, cf. \cite{Ilienko2}) null strings and
minimal null 2-surfaces. The task of formulating the conditions of
minimality for a null 2-surface in 4D~Minkowski spacetime seems to
be a rather difficult one. The standard approach of the classical
geometry of surfaces in the Riemannian space, which uses a
suitable variational principle, fails in this case. The problem
lies in the degenerate property of the induced metric (the first
fundamental form) of such surfaces and, therefore, one cannot
easily construct an analogue of area element like in the case of
time-like 2-surfaces.

Nevertheless, it is possible to formulate the minimality
conditions for a null two-surface in 4D~Minkowski spacetime. In
order to find them, a limiting procedure which takes the tangent
space to a space-like 2-surface element to that of a null
2-surface element was built in \cite{Disser}.

The conditions of minimality for a space-like 2-surface in
4D~Minkowski spacetime can be formulated in the same way as those
for a 2-surface in the ordinary Riemannian geometry. This has its
origin in the non-degenerate property of the first fundamental
form of a space-like 2-surface. In particular, the conditions of
minimality can be given by the requirement of vanishing of the
relevant mean curvatures. The mean curvatures are calculated by
taking a trace of the corresponding second fundamental forms.

Taking the limit of the well-defined minimality conditions for the
space-like 2-surfaces with the aid of that procedure, one finds
that minimal 2-surfaces admit a one parameter family of null
geodesics.

The geometry of the minimal null 2-surfaces depends on whether the
corresponding line of striction is a null or space-like curve. In
the former case the minimal null 2-surface is (locally)
developable and the null geodesics of the congruence are strongly
incident; in the latter case the null generators of the 2-surface
present an example of weakly incident light rays as has been
discussed recently by R.~Penrose \cite{Penrose_97}.

These results exhibit unusual features connected to the indefinite
nature of the Lorentz norm in 4D~Minkowski (and curved) spacetime.

\section{ACTION FOR STRINGS AND SPA\-CE-LI\-KE 2-SURFACES}

The method employed for obtaining the action functional
(\ref{Action}) of a null string can be in principle used for
designing an action for strings and a variational principle for
space-like 2-surfaces. The idea is to take a simple bivector
\begin{equation}
p_{ab}{}^*p{}^{ab}=\hphantom{-}0%
\label{Simple}
\end{equation}
and impose one of the conditions
\begin{eqnarray}
p_{ab}p{}^{ab}\!\! & = &\!\!\!\!\! \hphantom{-}1, \nonumber \\
p_{ab}p{}^{ab}\!\! & = &\!\!\!\!\! -1.%
\label{SS-L2-S}
\end{eqnarray}
The first condition in (\ref{SS-L2-S}) singles out the string
while the second corresponds to a space-like 2-surface. It is easy
to see that such a procedure uniquely fixes the symmetric second
rank spin-tensor in the standard decomposition of an
skew-symmetric 4D~Minkowski spacetime tensor
\begin{equation}
p{}_{ab}(x)=\phi{}_{AB}(x)\varepsilon{}_{A'B'} +
\bar{\phi}{}_{A'B'}(x)\varepsilon{}_{AB}.
\end{equation}
Then, the variational principle
\begin{equation}
S = \frac{1}{2!}\int p_{ab}(x)dx^a \wedge dx^b
\end{equation}
defines a 2-surface subject to the differential constraint
(\ref{DiffCon}). Now, one hopes that the use of spinor
decomposition for $p_{ab}(x)$ consistent with the algebraic
constraints (\ref{Simple}) and either of (\ref{SS-L2-S}) would
provide equations of motion, which automatically incorporate the
differential constraint. This assertion is supported by the
success of this procedure for the null bivectors outlined in the
current contribution. It may also be possible to derive the
analogues of the evolution equation (\ref{EvEqn}) for generic
(interacting) strings in 4D~Minkowski spacetime and curved
spacetimes of general relativity where exist explicit spinor
constructions.

In the same way one could build twistor action functionals in the
both cases and find corresponding objects on the null twistor
space for generic time-like and space-like 2-surfaces of
4D~Minkowski spacetime. This would accomplish the task of finding
of twistor description for 2-surfaces.

\section{ACKNOWLEDGEMENTS}

Thanks go to A.A.~Zheltukhin for drawing my attention to the
article \cite{Gusev-Zheltukhin}. I am very grateful to R.~Penrose
and Yu.P.~Stepanovskii for their interest in this work.


\begin{thebibliography}{99}
\bibitem{Schild}   A.~Schild, Phys. Rev.~D 16 (1977) 1722.
\bibitem{Stachel}  J.~Stachel, Phys. Rev.~D 21 (1980) 2182.
\bibitem{Gusev-Zheltukhin} O.E.~Gusev and A.A.~Zheltukhin,
                   JETP Lett. 64 (1996) 487.
\bibitem{Schouten} J.A.~Schouten, Ricci--Calculus,
                   Springer-Ver\-lag, 1954.
\bibitem{SST}      R.~Penrose and W.~Rindler, Spinors and
                   space-time, CUP, 1984.
\bibitem{Volkov:1988vf}
                   D.V.~Volkov and A.A.~Zheltukhin, JETP Lett.
                   48 (1988) 63.
\bibitem{Sorokin:1989zi}
                   D.P.~Sorokin, V.I.~Tkach and D.V.~Volkov, Mod.
                   Phys. Lett.~A 4 (1989) 901.
\bibitem{Sorokin:1989nj}
                   D.P.~Sorokin, V.I.~Tkach, D.V.~Volkov and
                   A.A.~Zhel\-tu\-khin, Phys. Lett.~B 216 (1989) 302.
\bibitem{Disser}   K.~Ilyenko, Twistor description of
                   null strings, DPhil Thesis, University of Oxford,
                   United Kingdom, 1999.
\bibitem{Bandos-Zheltukhin2} I.A.~Bandos and A.A.~Zheltukhin,
                   Theor. Math. Phys. 88 (1992) 925.
\bibitem{Hughston-Shaw} L.P.~Hughston and W.T.~Shaw, LMS Lecture Notes 156
                   (1990) 218.
\bibitem{Ilienko2} K.~Ilienko and A.A.~Zheltukhin, Class. Quantum Grav.
                   16 (1999) 383.
\bibitem{Penrose_97} R.~Penrose, Class. Quantum Grav. 14 (1997) A299.
\end{thebibliography}
\end{document}